\documentclass[12pt,preprint]{aastex}
\usepackage{epsfig}
\usepackage{amsmath}

\begin{document}
\title{Variability Tests for Intrinsic Absorption Lines in Quasar Spectra}

\author{Desika Narayanan\altaffilmark{1,2}, Fred
Hamann\altaffilmark{1}, Tom Barlow\altaffilmark{3},
E.M. Burbidge\altaffilmark{4}, Ross D.  Cohen\altaffilmark{4}, Vesa
Junkkarinen\altaffilmark{4}, Ron Lyons\altaffilmark{5}}
\altaffiltext{1}{Department of Astronomy, University of Florida 211
Bryant Space Science Center, Gainesville, FL 32611-2055, \\E-mail:
desika@astro.ufl.edu, hamann@astro.ufl.edu} 
\altaffiltext{2}{Current
Address: Steward Observatory, University of
Arizona, 933 N. Cherry Avenue, Tucson, AZ, 85721}
\altaffiltext{3}{Infrared Processing and Analysis Center, Caltech}
\altaffiltext{4}{Center for Astrophysics and Space Sciences,
University of California, San Diego} \altaffiltext{5}{PMB 353, 132
N. El Camino Real, Encinitas, CA 92024-2801}






\begin{abstract}
Quasar spectra have a variety of absorption lines whose origins range
from energetic winds expelled from the central engines to unrelated,
intergalactic clouds. We present multi-epoch, medium resolution
spectra of eight quasars at $z$ $\sim$ 2 that  have narrow
``associated'' absorption lines (AALs, within $\pm$5000 km s$^{-1}$
of the emission redshift).  Two of these quasars were also known
previously to have high-velocity mini-broad absorption lines
(mini-BALs).  We use these data, spanning $\sim$17 years in the
observed frame with two to four observations per object, to search
for line strength variations as an identifier of absorption that
occurs physically near (``intrinsic'' to) the central AGN.

Our main results are the following: Two out of the eight quasars with
narrow AALs exhibit variable AAL strengths. Two out of two quasars
with high-velocity mini-BALs exhibit variable mini-BAL strengths. We
also marginally detect variability in a high-velocity narrow
absorption line (NAL) system, blueshifted $\sim$32,900 km
$\rm{s}^{-1}$ with respect to the emission lines. No other absorption
lines in these quasars appeared to vary. The outflow velocities of the
variable AALs are 3140 km s$^{-1}$ and 1490 km s$^{-1}$. The two
mini-BALs identify much higher velocity outflows of $\sim$28,400 km
s$^{-1}$ and $\sim$52,000 km s$^{-1}$. Our temporal sampling yields
upper limits on the variation time scales from 0.28 to 6.1 years in
the quasar rest frames. The corresponding minimum electron densities
in the variable absorbers, based on the recombination time scale, are
$\sim$40,000 cm$^{-3}$ to $\sim$1900 cm$^{-3}$. The maximum distances
of the absorbers from the continuum source, assuming photoionization
with no spectral shielding, range from $\sim$1.8 kpc to $\sim$7 kpc.

\end{abstract}

\keywords{line: formation quasars: absorption lines, general}
\section{Introduction}

The absorption lines in quasar spectra can place important
constraints on the basic properties of quasar environments, such as
the outflow velocities, column densities, mass loss rates, and
elemental abundances. Additionally, if quasars reside
in the nuclei of massive galaxies, then we can use the chemical
abundances to probe indirectly the extent and epoch of star formation
in young galaxies (Schneider 1998, Hamann \& Ferland 1999).

Quasar absorption lines can be divided into three categories based on
the line widths: broad absorption lines (BALs), narrow absorption
lines (NALs), and intermediate mini-broad absorption lines
(mini-BALs). The divisions between these categories are
arbitrary. Classic BALs typically have full widths at half minimum
(FWHMs) of order 10,000 km s$^{-1}$, but lines as narrow as 2000 to
3000 km s$^{-1}$ are often still considered BALs (Weymann et al.
1991). A useful working definition of the NALs is that they have FWHMs
smaller than the velocity separation of major absorption doublets
(e.g., 500 km s$^{-1}$ for C IV $\lambda\lambda$1548,1551, or 960 km
s$^{-1}$ for N V $\lambda\lambda$1239,1243), but often these lines are
much narrower. Mini-BALs have widths intermediate between the NALs and
BALs.

BALs appear in about 10\% to 15\% of optically selected quasars
(Weymann et al. 1991). They appear at blueshifted velocities
(relative to the emission redshift) ranging from near 0 km s$^{-1}$
to $>$30,000 km s$^{-1}$, and they clearly form in high velocity
outflows from the quasar engines. Mini-BALs appear to be less common,
although to our knowledge no one has yet done a quantitative
inventory. Nonetheless, mini-BALs appear at the same range of
blueshifted velocities as the BALs and it is thought that they too
form in quasar winds (Hamann et al. 1997a, Jannuzi et al. 2003). 
Even though they may have complex profiles, high resolution spectra 
show that both BAL and mini-BAL profiles are ``smooth'' compared to 
thermal line widths, and therefore these lines are not simply blends 
of many NALs (Barlow \& Junkkarinen 1994, Hamann et al. 1997a, 
Junkkarinen et al. 2001, Hamann et al. 2003).

NALs appear at a wide range of velocity shifts. They are further
classified as associated absorption lines (AALs) if the absorption
redshift, $z_{\rm{abs}}$, is within $\pm$5000 km s$^{-1}$ of the
emission line redshift ($z_{\rm{abs}} \approx z_{\rm{em}}$; Weymann et
al. 1979, Foltz et al. 1986, Anderson et al. 1987, Foltz et al. 1988).
A significant fraction of AALs are believed to have a physical
relationship with the quasars, based on statistical correlations
between the occurrence of AALs and quasar properties (see also
Aldcroft et al. 1994, Richards et al. 1999). However, in general, AALs
and other NALs can form in a variety of locations, such as
cosmologically intervening clouds, galaxies that are unrelated to the
quasar, and clouds that are physically associated with, or perhaps
ejected from, the quasar (Weymann et al. 1979). NALs must therefore be
examined individually to determine if they are intrinsic to the
quasar. Several diagnostics have been proposed for this purpose,
including i) line strength variations over time, ii) profiles that are
smooth and broad compared to thermal line widths, iii) partial line of
sight coverage of the background emission source, and iv) high
densities based on excited-state absorption lines (e.g., Barlow \&
Sargent 1997, Barlow, Hamann \& Sargent 1997, Hamann et al. 1997a and
1997b).

Line strength variations can be caused by bulk motions across the line
of sight or by changes in the ionization state of the gas. A change in
the radial velocity of the absorbing material could also produce a
shift in the wavelength (redshift) of the measured lines (although
this type of shift appears to be extremely rare, Gabel et al. 2003,
this paper). In either case, variations over short time scales are
incompatible with absorption in large intergalactic clouds. A firm
lower limit on the size of intergalactic HI-absorbing clouds is
approximately a few kpc (Foltz et al. 1984, McGill 1990, Bechtold \&
Yee 1995, Rauch 1998), which is similar to recent direct estimates of
the sizes of intergalactic C~IV-absorbing clouds (Tzanavaris et
al. 2003).  Assuming the clouds are uniform and do not have sharp
edges, the time scale for absorption line strength variations
occurring via bulk motions will be roughly the time needed for this
minimum cloud to cross our line of sight. If the clouds have maximum
transverse velocities of $\sim$1000 km s$^{-1}$ (comparable to the
line-of-sight velocity dispersion in massive galaxy clusters), then in
the $\sim$30 years spanned by modern observations the cloud would
travel just $\sim$$3\times 10^{-5}$ kpc, roughly five orders of
magnitude less than its diameter. Variability due to changes in the
ionization state is also problematic for intergalactic absorbing
clouds. Changes in the ionization state are nominally limited by the
time scale for recombination, e.g., if the cloud is in ionization
equilibrium. Recombination on time scales of $\lesssim$30 years
requires densities of $n_e \gtrsim 100$ cm$^{-3}$ (see \S5.1 below),
which is exceedingly high compared to the densities $n_e\lesssim
10^{-3}$ cm$^{-3}$ expected in intergalactic H I clouds
(Miralda-Escude et al. 1996, Rauch 1998).  The variation time scales
might be shorter for lines forming in the (relatively) dense
interstellar medium of intervening galaxies, but these absorption
systems should be recognizable, e.g., via damping wings in
Ly$\alpha$. Therefore, line strength variations in NAL systems that do
not include damped Ly$\alpha$ strongly suggest that the absorption is
intrinsic to the quasar.

We are involved in a multi-faceted program to identify intrinsic NALs
and use them to place constraints on the physical properties of quasar
environments. In this paper, we examine multi-epoch rest frame UV
spectra of eight redshift $\sim$2 quasars known to have AALs and
mini-BALs. We find variability in 2 out of 8 AALs and 2 out of 2
mini-BALs.  We use these results to constrain the densities and
locations of the absorbing gas, and briefly discuss the implications
for quasar wind models.

\section{Observations and Data Reductions}

We obtained spectra of eight quasars at the Shane 3m telescope at the
University of California Observatories (UCO) Lick Observatory as well
as at the Smithsonian Institution and University of Arizona Multiple
Mirror Telescope (MMT) Observatory. Additional spectra obtained at
the Palomar Observatory were generously provided to us in digital
form by C. C. Steidel, first published in Sargent et al. (1988,
hereafter SBS88). Table 1 shows a log of the different observing
runs, including the dates, approximate wavelength ranges, and
resolutions. The data span $\sim$17 years ($\sim$6 years in
the quasar rest frames) with two to four observations per quasar.

In Table 2 we list the objects observed, their emission redshifts
($z_{\rm{em}}$, from Hewitt \& Burbidge 1993), the absorption
redshifts ($z_{\rm{abs}}$) and velocity shifts (relative to
$z_{\rm{em}}$) of their AALs, mini-BALs, or, in the case of
Q0151$+$048 only, a high-velocity NAL system that appeared to vary.
All of the AAL and NAL redshifts are from SBS88, except for PG0935+417
(from Hamann et al. 1997a) and Q0848+163 (see \S4 below). Note that
the AAL redshift given for PG0935+417 is an approximate average of a
complex of lines that is not resolved in our Lick spectra. Please see
SBS88 for a more complete list of absorption lines in the other
quasars. Table 2 also lists the observation numbers (from Table 1),
and a note on the variability of each system. The variabilities are
discussed further in \S4 below.

We chose the quasars for this study because they are bright, they are
known to have AALs, and their redshifts allow ground-based
observations of strong UV absorption lines, such as Ly$\alpha$
$\lambda$1216, N V $\lambda\lambda$1239,1243, Si IV
$\lambda\lambda$1394,1403, C IV $\lambda\lambda$1548,1551, and
potentially others in this wavelength range (see SBS88 and Junkkarinen
et al. 1991).

We reduced the data using standard techniques with the
IRAF\footnote{The Image Reduction and Analysis Facility (IRAF) is
distributed by the National Optical Astronomy Observatories, which is
operated by the Association of Universities for Research in
Astronomy, Inc. under cooperative agreement with the National Science
Foundation.} software. We bias subtracted the data and then divided
the raw two dimensional images by flat fields. Wavelength calibration
was achieved using  internal lamp spectra. We flux calibrated the
spectra using  standard stars measured the same night. Because the
weather was sporadically  cloudy, we used the flux calibrations
primarily to recover an accurate spectral shape, rather than absolute
fluxes. Finally, we added together spectra obtained at the same
wavelengths on the same dates to increase the signal to noise ratio.
The SBS88 spectra have a higher resolution than our Lick spectra; we
therefore smoothed the SBS88 data using a boxcar routine  to match
the resolution of  the Lick spectra and facilitate comparisons.

\section{Results}

We compared the spectra from the different observations of each
quasar to look for variations in the absorption line strengths. Every
absorption line within our wavelength coverage was examined, except
those in the Ly$\alpha$ forest. Table 2 presents the main results.
Two out of the eight quasars showed AAL variability. Two out of two
mini-BALs varied, and one quasar showed probable variations in a
high-velocity C IV NAL. Notes on the variable systems are provided in
\S4 below. Spectra showing the variations are also plotted in Figures
1--4.  The wavelength ranges in these plots were chosen to show only
the specific NALs or mini-BALs that varied. The peak flux in the
plotted wavelength range of the reference spectrum (solid curve) is
normalized to unity in all cases. Other spectra are over-plotted and
scaled to match approximately the reference spectrum in the continuum
near the lines of interest.

Our ability to detect line variability depends on the line strength
and the signal to noise ratio of the spectra. For the six AALs that
did not vary, we estimate upper limits on the changes, by inspection,
of the strongest measured lines (generally C IV and N V). In
particular, we estimate that the AAL equivalent widths varied by
$\lesssim$20 \% in Q0207$-$003, Q0348$+$061, PG0935$+$417, and
Q0958$+$551, and $\lesssim$15 \% in Q1159$+$123.

\section{Notes on Variable Sources}

\subsection{Q0151$+$048}

Figure 1 shows four epochs of observations for Q0151+048. We measured
clear variations in a C~IV mini-BAL at $z_{\rm{abs}}=1.6581$ and
tentative changes in a high-velocity C~IV NAL at
$z_{\rm{abs}}=1.6189$. No other lines are detected at these
redshifts. SBS88 report that both components of the narrow C~IV
doublet at $z_{\rm{abs}}=1.6189$ are blended with Si~II $\lambda$1526
absorption (see Fig. 1). However, much higher resolution Keck spectra
(Hamann et al. 2003) obtained in December 1996 provide no evidence for
these Si~II lines. In particular, we do not detect other low
ionization lines such as FeIV $\lambda$1608 or CII
$\lambda$1334. Also, the redshifts and profiles of the individual C~IV
doublet lines at $z_{\rm{abs}}=1.6189$ match each other very well,
indicating that all of the absorption is due to C~IV (at least for the
epoch December 1996). The data shown in Figure 1 suggest that these
lines varied. In particular, the equivalent width of the
short-wavelength component of the C~IV doublet is roughly twice as
strong in 1981 and 1996 compared to 1997. However, the two doublet
components seem to have varied by different amounts, which is
unphysical. Therefore, the variability in this system is
questionable. If this system is intrinsic, its displacement from the
emission lines implies an outflow velocity of $\sim$32,900 km
s$^{-1}$.

SBS88 describes the C IV mini-BAL in Q0151$+$048 as a blend of four
narrow C~IV absorption systems with redshifts ranging from
$z_{\rm{abs}}=1.6533$ to $z_{\rm{abs}}=1.6600$. We consider this
blend rather to be a mini-BAL, based on its appearance in the SBS88
spectrum and on the absence of discrete narrow components in our high
resolution Keck spectrum (Hamann et al. 2003). We measure the observed 
frame equivalent widths in this feature over the epochs monitored from 
0.38 \AA\ to 3.8 \AA , signifying a factor of $\sim$10 variation.
The FWHM of the mini-BAL (as observed in Oct. 1981) is
$\sim$1020 km s$^{-1}$, and its displacement from the emission
redshift is $-$28,400 km s$^{-1}$.

The AALs in this quasar did not appear to vary.

\subsection{PKS0424$-$131}
Figure 2 shows the variable N~V and C~IV AALs at
$z_{\rm{abs}}=2.1330$. The change in the N~V doublet is obvious. The
short wavelength component of the C~IV doublet is blended with an
unrelated line of FeII at $z_{\rm{abs}}=1.0352$. Ly$\alpha$ in this
AAL system is not clearly detected because of blending with another
unrelated line. We also do not detect Si~IV absorption at the AAL
redshift. The equivalent widths of the N~V and C~IV AALs varied by
55\% and 17\%, respectively, between the two observations.

Petitjean et al. (1994) published high resolution (0.3 \AA) spectra of
PKS0424$-$131 obtained in 1992. They report a total observed frame
equivalent width in the N~V doublet of 1.62 \AA , compared to 3.57
\AA\ and 1.72 \AA\ in our data from 1981 and 1997, respectively.
Petitjean et al. (1994) also measured an equivalent width of 0.31 \AA\
in the unblended long-wavelength component of the C IV doublet,
compared to our equivalent width measurements of 3.9 \AA\ and 0.36
\AA\ in our 1981 and 1997 spectra. These comparisons suggest that the
largest variations in these AALs occurred between the 1981 SBS88 and
1992 Petitjean et al.  observations.

\subsection{Q0848$+$163}
Higher resolution spectra (Hamann et al. 2003) confirm that there are
several AAL systems in this quasar. The two systems that SBS88
identified at $z_{\rm{abs}}\approx 1.916$ and $z_{\rm{abs}}\approx
1.917$ are blended together in our Lick spectra. We will therefore
refer to them hereafter as one system at $z_{\rm{abs}}$ = 1.9165.
There is another AAL system at $z_{\rm{abs}}$ = 1.9105 that SBS88 did
not identify, but that is unmistakable in our high resolution Keck
spectrum (Hamann et al. 2003).

Figure 3 provides some evidence for variability in these AAL
systems\footnote{The spectral region covering the CIV AALs in 0848+163
was measured and reported by SBS88.  However, these data were not
available for the present study, and are therefore excluded from
Figure 3.}.  Comparisons involving the February 1983 spectrum are
ambiguous because the underlying emission lines varied substantially
between that epoch and the more recent Lick observations. However, the
two 1997 observations suggest that similar variations occurred in the
N~V and C~IV doublets. The evidence for variability is perhaps
stronger for the system at $z_{\rm{abs}}$ = 1.9105, where the
equivalent width in the unblended short-wavelength component of the
N~V doublet declined by approximately 50\%. The AALs at
$z_{\rm{abs}}=1.9165$ might also have varied in 1997, but the changes
coinciding with the short-wavelength doublet components in this system
might be due at least in part to the variability at
$z_{\rm{abs}}=1.9105$. The variability in the accompanying Ly$\alpha$
lines is masked by blending with Ly$\alpha$ forest lines. We do not
detect the Si~IV (or other) lines in these systems.

\subsection{PG0935$+$417}
Figure 4 shows distinct variation in a C~IV mini-BAL that is
blueshifted by $\sim$52,000 km s$^{-1}$ with respect to the emission
lines (see also Hamann et al. 1997a). The identification of this
feature with highly blueshifted C~IV is confirmed by the appearance of
N~V and O~VI absorption at the same velocity (Rodriguez Hidalgo \&
Hamann 2003). This is the second highest observed velocity for a
mini-BAL, at nearly $\sim$0.2$c$, after PG2302$+$029 (Jannuzi et
al. 2003).  High-resolution Keck observations of PG0935$+$417 (Hamann
et al.  1997a) show that the mini-BAL remains smooth down to a
resolution of 7 km s$^{-1}$. This mini-BAL is therefore truly a broad
line with a continuous range of absorption velocities, and not a blend
of many NALs. The most dramatic variation occurred between January
1993 and March 1996 (Figure 4), where the observed frame equivalent
width in the mini-BAL increased from $\sim$5.12 \AA\ in 1993 to
$\sim$6.14 \AA\ in 1996 ($\sim$19\%). However, the change is most
obvious in the shape and centroid of the mini-BAL profile.  The
centroid of the line moved from $\sim$3917 \AA\ to $\sim$3865 \AA\
(corresponding to a shift of $\sim$4000 km s$^{-1}$) between 1993 and
1996. The FWHM of the mini-BAL was $\sim$1885 km s$^{-1}$ in 1993 and
$\sim$1320 km s$^{-1}$ on the other dates.

The complex of AALs in this quasar did not vary (see also Hamann et
al. 2003).

\section{Analysis}
The line variability time scales place constraints on the physical
properties of the absorbing clouds. Here we assume the variations
were caused by changes in the ionization state, which leads to
estimates of the minimum electron density and maximum distance from
the continuum source (for a photoionized plasma). The results are
listed in Table 3, where $t_{\rm{var}}$ is the smallest observed
variability time in the quasar rest frame, $n_{\rm{e}}$ is the
electron density, and $R$ is the distance from the continuum source.
Note that the values of $n_{\rm{e}}$ and $R$ are limits because we
measure only upper bounds on the variability times.

\subsection{Minimum Electron Densities}

We assume for simplicity that the gas is in ionization equilibrium
and the ions we measure are the dominant ionization stages. In that
case, the variability time is limited by the recombination time given
by,
\begin{equation}
n_e \ \approx \ \frac{1}{\alpha_{i-1} t^{\rm{recom}}_{i-1}}
\end{equation}
where $\alpha_{i-1}$ is the rate coefficient for recombination from
the observed ion stage $i$ to the next lower stage $i-1$, and
$t^{\rm{recom}}_{i-1}$ is the corresponding recombination time (see
Hamann et al. 1997b). Arnaud and Rothenflugh (1986) give recombination
rates for C IV $\rightarrow$ CIII and N V $\rightarrow$ NIV as 2.8
$\times$ $10^{-12}$ cm$^{3}$ s$^{-1}$ and 5.5 $\times$ $10^{-12}$
cm$^{3}$ s$^{-1}$, respectively. These values, together with a nominal
temperature of 20,000 K (Hamann et al. 1995) and the maximum
recombination times set by our observations, give the minimum electron
densities presented in Table 3. Because of the particular values of
the recombination rates, the calculated minimum electron densities for
N V are consistently a factor of $\sim$2 less than those for C
IV. Thus, we list constraints on the minimum electron densities in
Table 3 based only on the C IV recombination rates.

\subsection{Maximum Distances from Continuum Source}

We derive the maximum distance between the absorbing clouds and
the continuum source by assuming the gas is in photoionization
equilibrium with an ionization parameter given by,
\begin{equation}
U \ = \ \frac{1}{4 \pi R^{2} n_{H} \rm{c}}
\int_{0}^{\lambda_{\rm{LL}}}
\frac{\lambda L_{\lambda}}{hc} \rm{d} \lambda
\end{equation}
where $n_{H} \approx n_{e}$ is the hydrogen density,
$\lambda_{\rm{LL}} = 912$ \AA\ is the wavelength at the Lyman limit,
and $L_{\lambda}$ is the quasar luminosity distribution. We assume
all of the quasars in our sample have the same continuum shape
characterized by a segmented power law, $L_{\lambda}
\sim\ \lambda^{\alpha}$, where $\alpha = -1.6$ from 1000 \AA\ to
100,000 \AA\ , $-$0.4 from 10 \AA\ to 1,000 \AA\ , and $-$1.1 from
0.1 \AA\ to 10 \AA\ (Zheng et al. 1997, Telfer et al. 2002). We also
assume that the absorbing clouds are not ``shielded'' from the
continuum emission source. Thus the continuum spectrum incident on
the absorbing clouds is diminished only by the geometric 1/$R^2$
dilution. Integrating over our adopted continuum shape in Equation 2
yields a convenient expression:
\begin{equation}
U \ \approx \ 0.09 \frac{L_{47}}{n_{10}(R_{1})^{2}}
\end{equation}
where L$_{47}$ is the bolometric luminosity in units of 10$^{47}$ erg
s$^{-1}$, $n_{10}$ is the density in units of 10$^{10}$ cm$^{-3}$,
and $R_{1}$ is the distance from the continuum source in parsecs.
Note that for this continuum shape, the quasar bolometric luminosity
is $L_{\rm{bol}} \approx 4.4\,\lambda L_{\lambda}$ at $\lambda=1450$
\AA.

We derive bolometric luminosities for the quasars in our sample based
on the continuum shape given above, observed $B$ magnitudes from
Junkkarinen et al. (1991) and Hewitt \& Burbidge (1993), and a
cosmology with $H_{0}=72$ km s$^{-1}$ Mpc$^{-1}$, $\Omega_{M}=0.3$,
and $\Omega_{\Lambda}=0.7$. For PKS0424$-$131, we use a measured $V$
magnitude and a representative value of $B-V = 0.12$ (based on other
luminous quasars at this redshift, Hewitt \& Burbidge 1993). The
resulting values of $L_{47}$ are listed in Table 3. Finally, we adopt
an ionization parameter of U $\approx$ 0.02, which is approximately
optimal for the C IV and N V ions (e.g., Hamann et al. 1995, Hamann
1997a). Plugging into Equation 3 with the previously derived density
limits yields the upper limits on $R$ listed in Table 3.





\section{Discussion}
Two of the 8 quasars in our study showed variability in their AALs.
This fraction is similar to the 3 out of 15 AAL systems showing
variability in a sample of lower redshift quasars studied by Wise et
al. (2003). Note that these results provide only minimum variability
fractions because of the limited time sampling. The fraction of AALs
that are intrinsic based on variability is therefore $\ga$25\%. We
are now analyzing much higher resolution spectra of all of the
quasars in our sample (Hamann et al. 2003) to look for other evidence
of intrinsic absorption (\S1).

The mini-BALs appear to vary more often than the AALs. Our study showed
variations in 2 out of 2 systems. The only other mini-BALs tested for
variability, that we know of, are PG~2302+029 (Jannuzi et al. 2003)
and Q2343+125 (which is marginally a mini-BAL with FWHM $\approx$ 400
km s$^{-1}$, Hamann et al. 1997c). Both of those systems also varied
between just two observations.

The marginal detection of variability in the high-velocity NAL system
of Q0151+048 is surprising but not unprecedented. Richards et al.
(1999) argued that a significant fraction of NALs at blueshifted
velocities from 5000 to 75,000 km s$^{-1}$ are intrinsic to quasars,
based on statistical correlations between the appearance of these
lines and the quasar radio properties. The high-velocity NAL system
in Q0151+048 might be one specific example at a velocity shift of
$\sim$32,900 km s$^{-1}$.

It is interesting to note that, with one exception, the variable
absorption lines remained fixed in velocity.  The exception is the
mini-BAL in PG0935+417 whose changing centroid and FWHM was noted
previously by Hamann et al (1997a).  However, it is not clear that
these velocity changes represent real changes in the absorber
kinematics.  It could be that the apparent velocity changes are caused
by variations in the line strength (optical depth) across a fixed
range of absorption velocities.  A more likely example of a real
velocity shift involves an AAL system in the Seyfert I galaxy, NGC
3783 (Gabel et al. 2003).

All of the variable absorption lines in our sample appear at {\it
blueshifted} velocities, clearly indicating outflows from the
quasars. However, the wind geometry implied by these AALs and
mini-BALs must be different from the classic BALs. In particular, the
line widths are small compared to the wind speed. BALs typically have
$V$/FWHM of order unity (where $V$ is the outflow speed). In contrast,
the mini-BAL in Q0151+048 has $V$/FWHM $\approx$ 28, and the
high-velocity NAL in that quasar has $V$/FWHM $\approx$ 55. Another
intrinsic absorber in Q2343+125 (Hamann et al. 1997c) has $V$/FWHM
$\approx$ 60. Absorption lines like these with $V$/FWHM $\gg$ 1
probably do not form in continuous outflows that accelerate from rest
along our line of sight to the continuum source. If that were the
case, we should see absorption at all velocities from $V$ to 0,
unless, perhaps, there is a peculiar ionization structure that somehow
favors C IV and similar ions only at the very narrow range of observed
line velocities. A more likely possibility is that the observed lines
form in discrete ``blobs'' of gas that were ejected/accelerated
sometime prior to the observations. However, the nature of these blobs
as coherent entities is unclear. The large velocity dispersions
implied by the line widths (FWHM $>$ 1000 km s$^{-1}$ for the
mini-BALs) should quickly lead to a spatially extended
structure. Another possibility is that the flows intersect our line of
sight to the continuum source (and become observable in absorption)
only after they reach the observed high speeds (see also Hamann et
al. 1997c, Elvis 2000). Thus the acceleration occurs somewhere outside
of our sight line to the continuum source. In this case the flow could
be continuous -- simultaneously spanning a wide range of velocities
and radial distances, $R$, but the absorption lines sample only a
limited range of these parameters, depending on the specific geometry
and orientation.

One plausible geometry involves a wind emanating from the quasar
accretion disk (Murray et al. 1995, Proga, Stone \& Kallman 2000). If
the lines we measure form anywhere near the point of origin of these
winds, within several parsecs of the central black hole and possibly
much closer, then simple considerations based on Equation 3 suggest
that the actual gas densities are $n_e \gtrsim 10^9$
cm$^{-3}$. Unfortunately, the lower limits on $n_{\rm{e}}$ in Table 3
are not very constraining.  They are inconsistent with only the most
extreme lower densities and large radial distances ($\gtrsim$10 kpc)
derived for some AAL systems (based on excited-state absorption lines,
e.g., Morris et al. 1986, Tripp, Lu \& Savage 1996, Hamann et
al. 2001).

\section{Conclusions}

We observed eight AAL quasars to test for variability in their
absorption line strengths as an indicator of intrinsic absorption.
Two of these quasars were also known previously to have high-velocity
mini-BALs. In our limited time sampling of 2--4 observations per
quasar, we found variability in both of the mini-BALs, 2 out of 8 AAL
systems, and possibly one additional high-velocity NAL system. These
results agree with previous reports of frequent variability in
mini-BALs (Hamann et al. 1997a, Jannuzi et al. 2003), and with the
recent finding of AAL variability in 3 out of 15 quasars studied by
Wise et al. (2003). The short time scales over which the lines varied
in our study (sometimes $<$0.28 years in the quasar rest frame)
implies that the absorbers are dense, compact, and physically
associated with the quasars. We estimate minimum electron densities
(from the recombination time) ranging from $\sim$1900 cm$^{-3}$ to
$\sim$40,000 cm$^{-3}$. We conclude that the fraction of AALs that
are intrinsic based on variability is at least $\sim$25\%. The
fraction of mini-BALs that are intrinsic may be $\sim$100\%. The
outflow velocities implied by the intrinsic systems range from
$\sim$1500 km s$^{-1}$ to $\sim$52,000 km s$^{-1}$.

\acknowledgements We are grateful to C.C. Steidel for generously
providing the Palomar data.  This work was supported by NSF grant AST
99-84040. D.N. thanks Craig Warner for helpful discussions, and
acknowledges the University Scholars Research Program at the
University of Florida for financial support.

\clearpage

\begin{deluxetable}{ccccc}
\tabletypesize{\scriptsize}
\tablecaption{Observation Information}
\tablewidth{0pt}
\tablehead{
\colhead{Obs \#} &\colhead{Observatory} & \colhead{Date} & \colhead {$\lambda$ Range (\AA)}
 & \colhead{Res (\AA)}}
\startdata
1& Palomar& Nov 1981&3300-4800&0.8-1.5\\
2& Palomar& Feb 1983&3100-3700&0.8\\
3& Palomar& Nov 1983&3100-5000&1.5\\
4& Palomar& May 1984&4600-5800&1.5\\
5& Palomar& Oct 1984&4700-6000&1.5\\
6& Palomar& Apr 1985&4800-6600&0.8-2.2\\
7& Lick & Jan 1993 & 3250-5350 & 2.95\\
8& Lick & Mar 1996 & 3250-4600 & 2.95\\
9& MMT & Dec 1996 & 3400-4400 & 1.4\\
10& Lick & Feb 1997 & 3200-6000 & 2.95\\
11& Lick & Dec 1997 & 3100-6000 & 2.95\\
12& Lick & Jan 1999 & 3200-6000 & 2.95\\
\enddata
\end{deluxetable}

\newpage

\begin{deluxetable}{cccrccc}
\tablecaption{Absorption Line Properties}
\tabletypesize{\scriptsize}
\tablewidth{0pt}
\tablehead{
\colhead{QSO} & \colhead{$z_{\rm{em}}$} & \colhead{$z_{\rm{abs}}$} &
\colhead{$v$ (km s$^{-1}$)}& \colhead{Type}&
\colhead{Obs \#} & \colhead{ var?}
}
\startdata
Q0151$+$048 & 1.9232  &1.6189 &$-$32,900 & NAL&1,9,11,12 &yes?\\
         &         &1.6581 &$-$28,400 & mini-BAL&        &yes\\
         &         &1.9343  &$+$900      & AAL & &no\\
Q0207$-$003 & 2.849   &2.8871 &$+$2950 &AAL&5,11,12 &no \\
Q0348$+$061& 2.060 & 2.0237 &$-$3580 & AAL &3,12 & no \\
         &       & 2.0330  &$-$2660  &AAL&     &no  \\
PKS0424$-$131 & 2.166 & 2.1330  &$-$3140 &AAL&1,2,11 &yes \\
         &       & 2.1731  &$+$660 &AAL&       &no \\
Q0848$+$163 & 1.925 & 1.9105  &$-$1490 &AAL&1,10,11,12 &yes \\
         &       & 1.9165  &$-$870  &AAL&           &yes? \\
PG0935$+$417 & 1.966 & 1.490   &$-$52,000 &mini-BAL&7,8,11,12 &yes\\
 & & 1.938& $-$2780& AAL& & no\\
Q0958$+$551 & 1.751 & 1.7310  &$-$2190 &AAL      &3,12 &no \\
         &       & 1.7327  &$-$2000 & AAL        &     &no \\
Q1159$+$123 & 3.502 & 3.5265  &$+$1630 & AAL       &4,6,10 &no \\
\enddata
\end{deluxetable}

\newpage
\begin{deluxetable}{cccccc}
\tabletypesize{\scriptsize}
\tablecaption{Physical Properties}
\tablewidth{0pt}
\tablehead{
\colhead{Object}&
\colhead{$z_{\rm{abs}}$} &
\colhead{$t_{\rm{var}}$ (years)}&
\colhead{$n_{\rm{e}}$ (cm$^{-3}$)}&
\colhead{$L_{47}$}& \colhead{$R$ (pc)}
}
\startdata
Q0151$+$048&1.6189&0.40&$\geq$25,000& 1.9& $\leq$1800  \\
        &1.6581&5.67&$\geq$2000& --- & $\leq$6540\\
PKS0424$-$131&2.1330&3.37&$\geq$3400& 3.7& $\leq$7000\\
Q0848$+$163&1.9105&0.28&$\geq$40,000& 6.4& $\leq$2700\\
PG0935$+$417& 1.490 &1.27&$\geq$8900& 9.6& $\leq$7000\\
\enddata
\end{deluxetable}

\clearpage

\figcaption[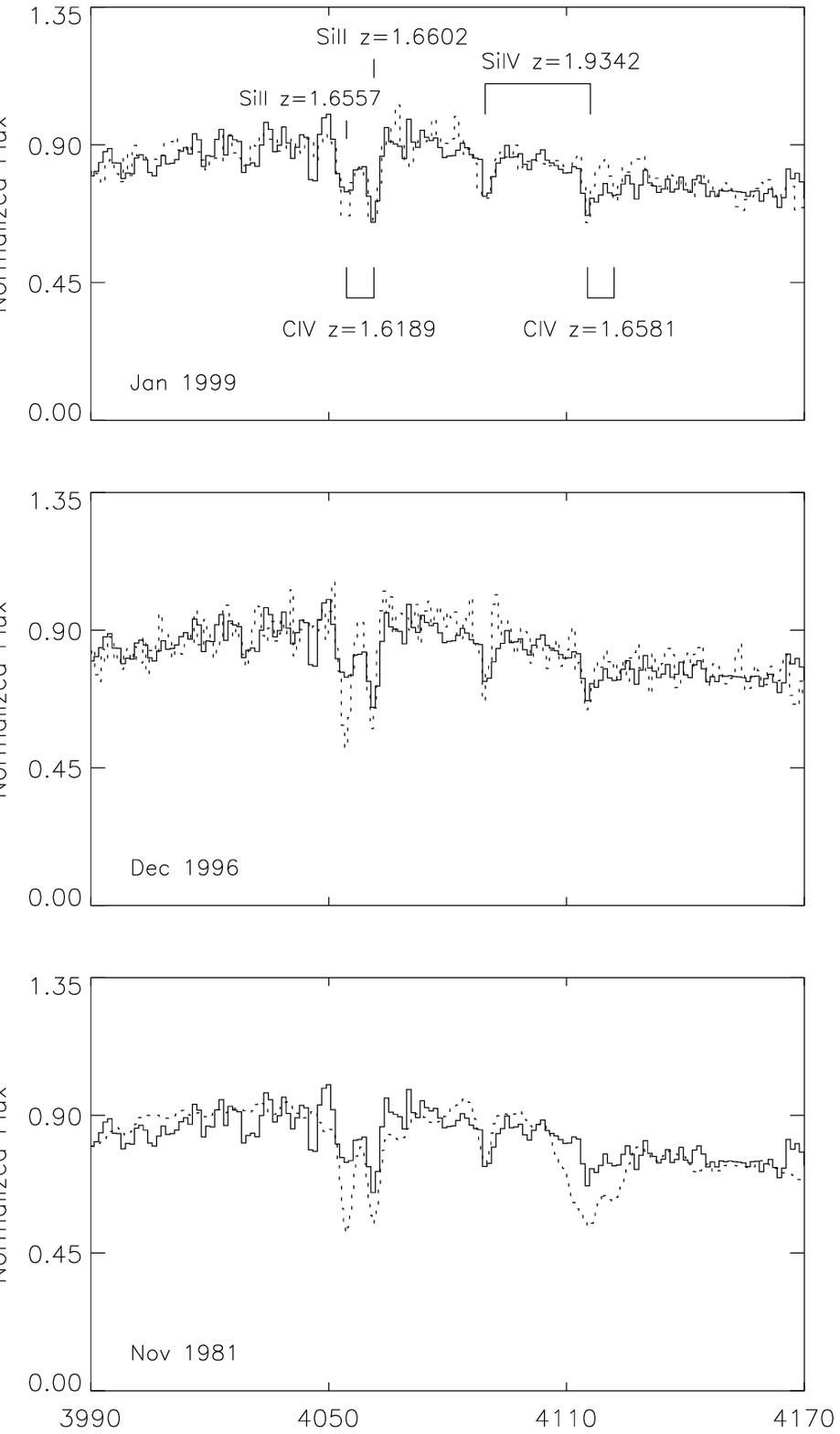]{
Q0151$+$048:  The solid curve in each panel is the normalized
December 1997 spectrum, while the dotted curves are other spectra
from the dates indicated. The C~IV mini-BAL at $z_{abs} = 1.6581$ and
the high-velocity C~IV NALs at $z_{abs}=1.6189$ are labeled below
the spectra in the top panel. Other unrelated lines are labeled
above. The two Si~II identifications are probably spurious, based on
much higher resolution spectra obtained in December 1996. The bottom
panel shows clearly the variability in the mini-BAL (moving at
$\sim$28,400 km s$^{-1}$). The apparent variation in the C~IV
NALs at $z_{\rm{abs}}$=1.6189 is suspect because one component of
this doublet changed more than the other (see \S4.1).
}

\figcaption[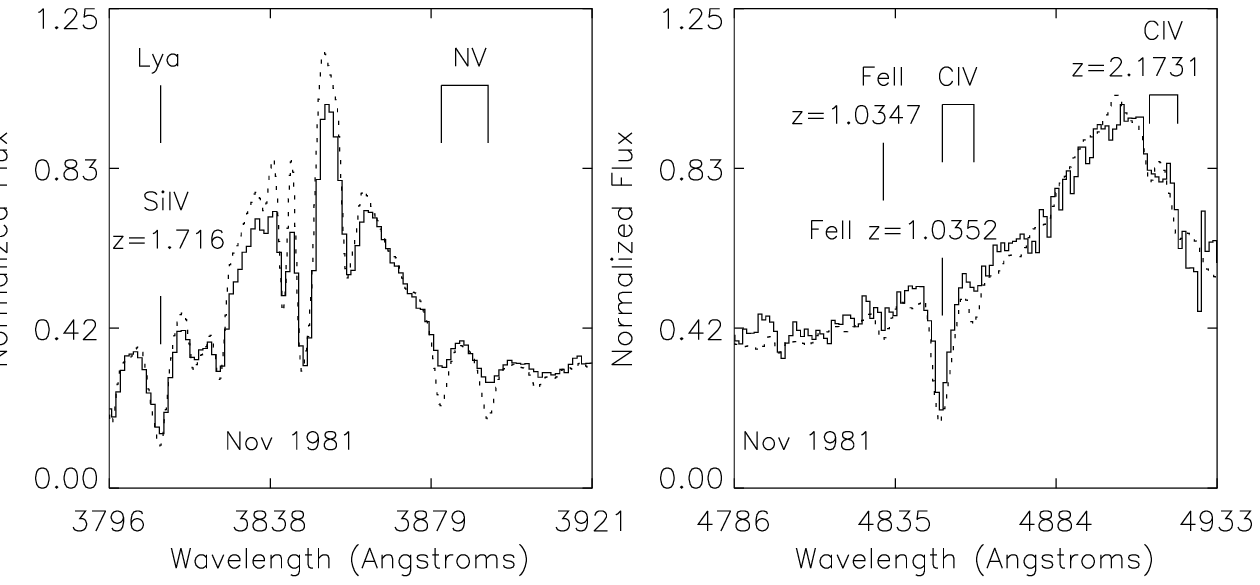]{
PKS0424$-$131. Normalized spectra showing variation in a
$z_{\rm{abs}}$=2.1330 AAL system. The Ly$\alpha$, N~V, and C~IV lines
in this variable AAL system are labeled without redshifts. Some other
unrelated lines are also labeled with redshifts. The solid curve in
both panels is the Dec. 1997 spectrum, while the dotted curves show
the Nov. 1981 observation. The wavelength range in both panels
corresponds to $\sim$10,000 km s$^{-1}$.}

\figcaption[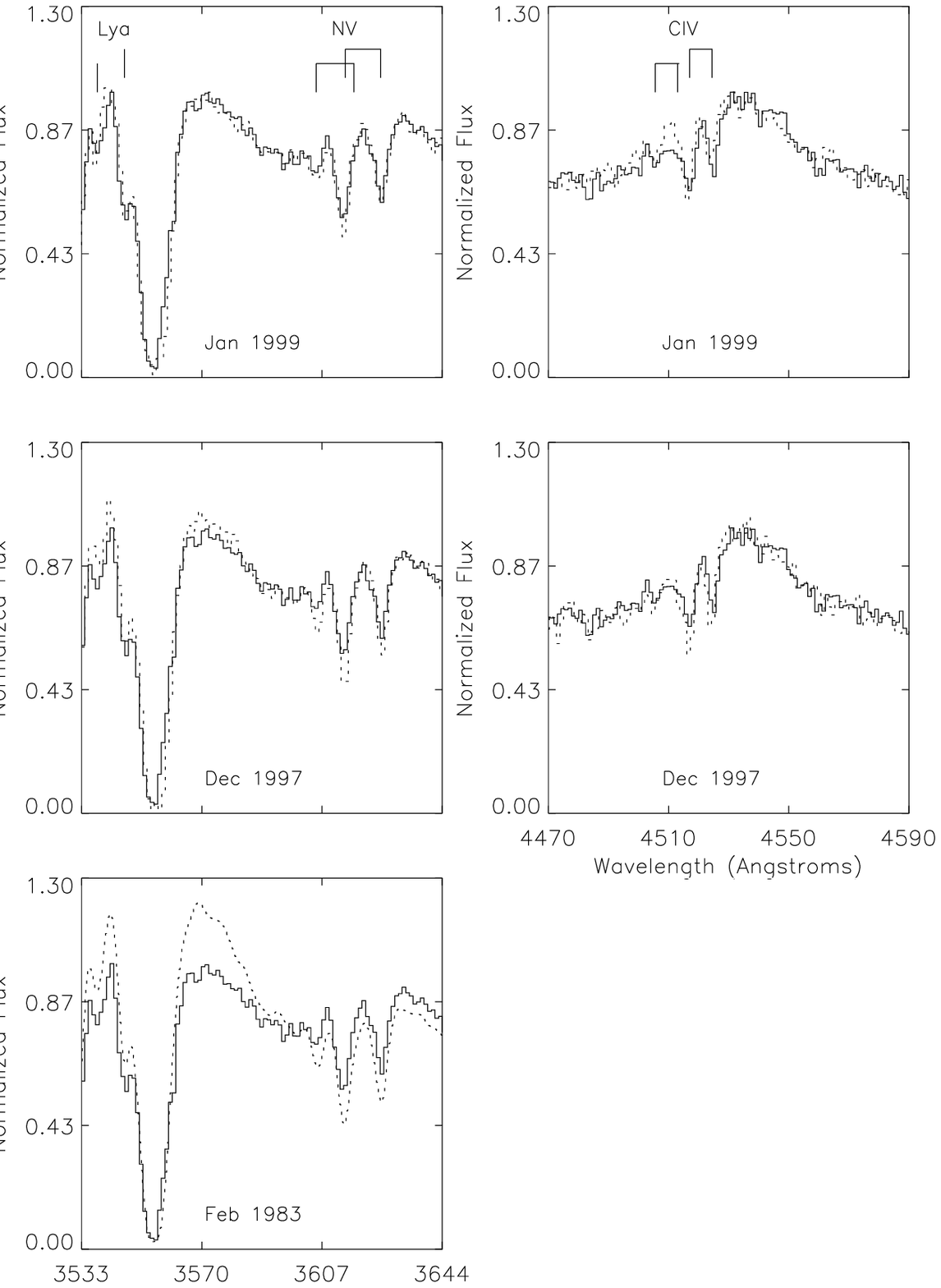]{ Q0848$+$163. Normalized spectra showing variable
AALs at $z_{\rm{abs}}$=1.9105 and $z_{\rm{abs}}$=1.9165. The red
components of the N~V and C~IV doublets at $z_{\rm{abs}}$=1.9105 are
blended with the blue components of these doublets at
$z_{\rm{abs}}$=1.9165.  The solid line is a Feb 1997
observation. Please note that, for the purpose of best showing the
variability, the fiducial plot in this figure is different from the
rest of the figures. Dates for the dotted spectra are given in each
panel. Unfortunately, we do not have Palomar data covering the C~IV
AALs.  The strong, unlabeled line at 3555 \AA\ is Ly$\alpha$ at
$z_{\rm{abs}}$=1.925.  The wavelength range in all panels corresponds
to $\sim$10,000 km s$^{-1}$.}

\figcaption[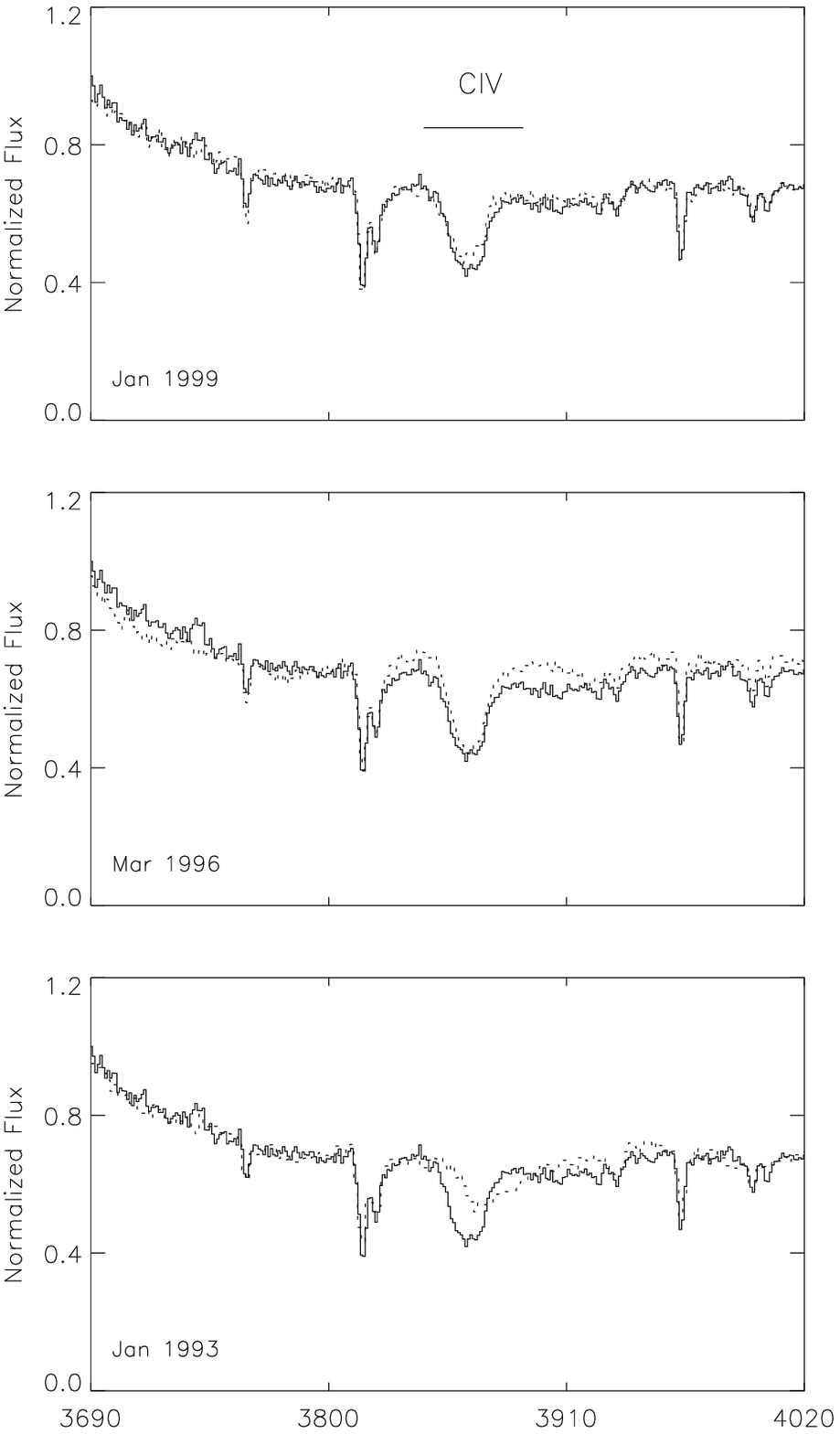]{ PG0935$+$417: The solid curve is the normalized
December 1997 spectrum. The horizontal bar at $\sim$3875 \AA\ in the
top panel marks the approximate extent of the C~IV mini-BAL. This
absorber is outflowing at nearly 52,000 km s$^{-1}$. Several unrelated
NALs are also present, but not labeled in this plot.}

\newpage
\epsscale{0.7}
\plotone{f1.eps}
\newpage
\epsscale{1}
\plotone{f2.eps}
\newpage
\epsscale{0.9}
\plotone{f3.eps}
\newpage
\epsscale{0.7}
\plotone{f4.eps}

\end{document}